\documentclass[a4paper,10pt,twoside]{cpc-hepnp}

\usepackage{multicol}
\usepackage{graphicx}
\usepackage{booktabs}
\usepackage{amssymb,bm,mathrsfs,bbm,amscd}
\usepackage[tbtags]{amsmath}
\usepackage{lastpage}
\usepackage{CJK}

\begin{document}

\fancyhead[co]{\footnotesize Li Hai-Bo, Zhu Shi-Hai: Mini-review of
rare charmonium decays at BESIII}

\footnotetext[0]{Received 1 December 2011}

\title{ Mini-review of rare charmonium decays at BESIII \thanks{Supported by National Natural Science
Foundation of China (11125525) }}

\author{%
     Li Hai-Bo\email{lihb@ihep.ac.cn}%
\quad Zhu Shi-Hai\email{zhush@ihep.ac.cn}%
 } \maketitle

\address{%
Institute of High Energy Physics, Chinese Academy of Sciences, Beijing 100049, China\\
}

\begin{abstract}
Recently, LHCb experiment announced a 3.5$\sigma$ evidence for
direct $CP$ violation in $D^0$ decay by looking at the difference
between $A_{CP}(D^0\rightarrow K^+K^-)$ and $A_{CP}(D^0\rightarrow
\pi^+\pi^-)$. This is the first evidence of $CP$ violation in charm
system, which may indicate new physics beyond the Standard Model.
Motivated by this measurement, we review rare processes in
charmonium decay, especially, the weak decay, $C$ or $P$ violated
decay and lepton flavor violated decays. In case the new physics
appears in charm sector, these rare decays of charmonium states will
provide opportunity to search for significant contributions from
physics beyond the Standard Model. With huge $J/\psi$ and $\psi(2S)$
samples in BESIII experiment, the rare decays may be feasible.

\end{abstract}



\begin{multicols}{2}

\section{Introduction}
 At present, two general trends can be distinguished in accelerator
particle physics. On one hand,  very high energy accelerators, for
example, the LHC,  provide the ability to explore physics at high
energy frontier. On the other hand, smaller experiments at lower
energies but with very high intensities and low backgrounds, for
example, the $B$ factories and BESIII , provide the capabilities for
performing precise tests and accurate determinations of many
parameters of the Standard Model (SM). Moreover, the close scrutiny
of rare processes may illuminate new physics in a complementary
fashion to high-energy colliders.

On November 14, 2011, LHCb experiment announced 3.5$\sigma$ evidence
for direct $CP$ violation in $D^0$ decay.  They looked at the
difference between $A_{CP}(D^0\rightarrow K^+K^-)$ and
$A_{CP}(D^0\rightarrow \pi^+\pi^-)$, and found that the $\Delta
A_{CP} = [-0.82\pm0.21(\mbox{stat.}) \pm
0.11(\mbox{sys.})]$\%~\cite{lhcb2011}. Since the SM predicted that
the $CP$ asymmetry in charm sector should be less that
0.1\%~\cite{grossman}, this is the first indication of new physics
at LHC. If the new physics could appear in the up quarks (charm and
top), it urgently needs confirmation. One possible way is to probe
the charmonium rare decays with huge $J/\psi$ and $\psi(2S)$ data
sample. The BESIII experiment can provide important tests of the SM,
with the accompanying possibility for uncovering new physics induced
deviations in charm sector, especially, the weak charmonium decay,
invisible decay, lepton flavor violated decays and so on.  In this
paper, we review the possibility of such kinds of rare charmonium
processes.


\section{Weak decays of charmonium}
\label{sec:weak_cc_decays}

\indent  The low-lying charmonium states, {\it i.e.} those below the
open-charm threshold, usually decay through intermediate photons or
gluons produced by the annihilation of the parent $c\bar{c}$ quark
pair. These OZI-violating but flavor-conserving decays result in
narrow natural widths of the $J/\psi$ and $\psi(2S)$ states. In the
SM framework, flavor-changing weak decays of these states are also
possible, although they are expected to have rather low branching
fractions. The huge $J/\psi$ data samples at BESIII will provide
opportunities to search for such rare decay processes, which in some
cases may be detectable, even at SM levels. The observation of an
anomalous production rate for single charmed mesons in $J/\psi$ or
$\psi(2S)$ decays at BESIII would be a hint of possible new physics,
either in underlying continuum processes via
flavor-changing-neutral-currents~\cite{lihb_neutral,lihb_sl_psi,verma-2009,wang-2008-semi,wang-2008,wang-2009,wang-2008-epj}
or in the decays of the $\psi$ resonances due to unexpected effects
of quark dynamics~\cite{lihb_weak,li-2009}.

\subsection{ Semileptonic decays of charmonium}

The inclusive branching fraction for $J/\psi$ weak decays via a
single quark (either the $c$ or the $\bar{c}$) had been estimated to
be $(2\sim4) \times 10^{-8}$ by simply using the $D^0$
lifetime~\cite{lihb_sl_psi}. Such a small branching fraction  makes
the observation of weak decays of the $J/\psi$ or $\psi(2S)$ quite
challenging, despite the expected cleanliness of the events.
However, BEPC-II, running at designed luminosity, will produce of
order  $10^{10}$ $J/\psi$ events per year of data taking, leading to
$ \cong 400$ weak decays for the predicted SM branching fraction.
The semi-leptonic decay of a $c\bar{c}~(1^{--})$ vector charmonium
state below the open-charm threshold is induced by the weak
quark-level transition $c \rightarrow q W^*$, where $W^*$ is a
virtual intermediate boson. Hence, the accessible exclusive
semi-leptonic channels are:
\begin{eqnarray}
\psi(nS) \rightarrow D_q l \nu, \label{eq:semi_c_allowed_1}
\end{eqnarray}
\begin{eqnarray}
\psi(nS) \rightarrow D_q^* l \nu, \label{eq:semi_c_allowed_2}
\end{eqnarray}
where $n = 1$~or~$2$, and $q$ can be either a $d$- or $s$-quark,
which corresponds to a $D^{\pm}$(Cabibbo-suppressed mode) or $D_s$
(Cabibbo-allowed mode) meson.
 Semi-leptonic weak decays of the
$J/\psi$ will offer several advantages over the purely hadronic ones
from both the experimental and theoretical points of view: the
prompt charged lepton $l = e, \mu$ can be used to tag the events,
removing a large fraction of conventional $\psi(nS)$ hadronic
decays. In addition, the missing energy due to the escaping neutrino
can be also exploited to remove backgrounds.  The identification of
the charm meson in the final state would provide an unambiguous
signature of the semi-leptonic weak decays of $\psi(nS)$. Meanwhile,
decays of the excited mesons $D^*_s$ and $D^{*\pm}$ produced in
reaction (\ref{eq:semi_c_allowed_2}) would provide useful additional
experimental handles.  In the lab system, the detectable photons
from the $D^{*\pm}_s \rightarrow D^{\pm}_s \gamma$, radiative
transition are  in the 90$\sim$200~MeV energy interval. These, and
the soft pion produced from $D^{*\pm}$ in $D^{*\pm} \rightarrow
D^{0} \pi^{\pm}$ decay, can provide powerful constraints to help
identify a $D_s$ or $D^0$ meson produced in the weak decay of a
charmonium state.

Many theoretical calculations have been done based on various QCD
frameworks, including, e.g., heavy quark spin symmetry
(HQSS)~\cite{lihb_sl_psi}, QCD sum rules~\cite{wang-2008-semi} and
covariant light-front quark model~\cite{wang-2008}. The predicted
branching fractions are in the range of $10^{-9} - 10^{-10}$ and
$10^{-10} - 10^{-11}$ for $J/\psi \rightarrow D^{(*)}_s l \nu$ and
$J/\psi \rightarrow D^{(*)} l \nu$, respectively. The branching
fractions for the sum of dominant semi-leptonic decay modes can
arrive at $5\times 10^{-9}$~\cite{wang-2008}, which are hopefully to
be detected at BESIII with $10 \times 10^9$ $J/\psi$ events
accumulated each year. The overall theoretical uncertainty is up to
$\sim 50\%$ due to the treatment of the nonperturbative QCD
dynamics. In Table~\ref{tab:semi_mode_cc}, the theoretical
predictions for $J/\psi$ semi-leptonic decays are summarized.
\end{multicols}

\begin{center}
  \tabcaption{\label{tab:semi_mode_cc}  The predicted branching fractions (in units of $
  10^{-10}$)
   for $J/\psi$ semi-leptonic
  weak decays. The transition form factors for $J/\psi \rightarrow D^{(*)}_{s,d}$ are estimated
  based on the ISGW model in Ref.~\cite{lihb_sl_psi}, the factorization scheme in Ref.~\cite{verma-2009},
  the QCD sum rules in Ref.~\cite{wang-2008-semi} and the covariant light-front quark model in Ref.~\cite{wang-2008}.
    Here, ${\cal BR}(J/\psi \rightarrow D^{*-}_{s,d} l^+ \nu) ={\cal BR}(J/\psi \rightarrow D^{*-}_{s,d} e^+ \nu_e)+
   {\cal BR}(J/\psi \rightarrow D^{*-}_{s,d} \mu^+ \nu_{\mu})$. Only the central values are cited from
   the theoretical predictions.  }
  \begin{tabular}{c|c|c|c|c}\hline
\hline decay mode  & Ref.~\cite{lihb_sl_psi}   &
Ref.\cite{verma-2009} & Ref.~\cite{wang-2008-semi} &
Ref.~\cite{wang-2008}
 \\ \hline

 $J/\psi \rightarrow D^-_s l^+ \nu+c.c.$   & 26 & 20 & 3.5
    &  11.2 \\ \hline
$J/\psi \rightarrow D^{*-}_s l^+ \nu+c.c.$ & 42 & 32 & 11
    & 34.7  \\ \hline
$J/\psi \rightarrow D^- l^+ \nu+c.c.$ & 1.4 & 1.2 & 0.14
    & 1.1 \\ \hline
$J/\psi \rightarrow D^{*-} l^+ \nu+c.c.$   & 2.3 & 2.0 & 0.73
    &  3.4 \\ \hline
\hline
  \end{tabular}
\end{center}

\begin{multicols}{2}

It is interesting to note that the ratio of Cabibbo-allowed decay to
Cabibbo-suppressed decay can be obtained cleanly since many
theoretical uncertainties cancel out in the ratio. The ratios
$R_{s/d} = \frac{{\cal BR}(J/\psi \rightarrow D^-_s l^+ \nu)}{{\cal
BR}(J/\psi \rightarrow D^{-} l^{+} \nu)}$ and $R^*_{s/d} =
\frac{{\cal BR}(J/\psi \rightarrow D^{*-}_s l^+ \nu)}{{\cal
BR}(J/\psi \rightarrow D^{*-} l^{+} \nu)}$ should be equal to
$\frac{|V_{cs}|^2}{|V_{cd}|^2 }\cong 18.4$ under the SU(3) flavor
symmetry limit~\cite{lihb_sl_psi}, where $V_{cs}= 0.973$ and
$V_{cd}= 0.2271$ denote the relevant Cabibbo-Kobayashi-Mashkawa
(CKM) mixing matrix elements. In Ref.~\cite{wang-2008-semi}, the
calculations based on the QCD sum rules show $R_{s/d} \cong 24.7$
and $R^*_{s/d} \cong 15.1$, which implies a large effect of the
SU(3) symmetry breaking.  Under the assumption of heavy-quark spin
symmetry (HQSS) and the non-recoil approximation~\cite{lihb_hqss_1,
lihb_hqss_2}, the ratio of $R_{V/P} = \frac{{\cal BR}(J/\psi
\rightarrow D^{*-}_s l^+ \nu)}{{\cal BR}(J/\psi \rightarrow D^-_s
l^{+} \nu)}$ is predicted to be 1.6~\cite{lihb_sl_psi}, while it is
calculated to be 3.1 based on the QCD sum rules in
Ref.~\cite{wang-2008-semi}, which indicates that high order
corrections are important.

Following the first proposal by Zhang~\cite{zhang-2001}, BESII has
measured the $J/\psi$ weak decays to be ${\cal BR}(J/\psi\rightarrow
D^-_s e^+\nu +c.c.) <3.6 \times 10^{-5}$ and ${\cal
BR}(J/\psi\rightarrow D^- e^+\nu +c.c.) <1.2 \times 10^{-5}$ at 90\%
confidence level based on $5.8 \times 10^{8}$ $J/\psi$ decay
events~\cite{bes2-2006}. The current experimental limits are much
higher than the predicted values. It will be important to search for
$J/\psi$ weak decays at the BESIII experiment, which will be
discussed in detail in Section 2.3.

The flavor-changing-neutral-current (FCNC) induced semi-leptonic
decays of $J/\psi\rightarrow \bar{D}^{(*)0}l^+l^-$ are studied in
Ref.~\cite{wang-2009} based on the QCD sum rules, and the decay
rates for $J/\psi\rightarrow \bar{D}^{0}l^+l^-$ and
$J/\psi\rightarrow \bar{D}^{*0}l^+l^-$ are predicted to be
$10^{-14}$ and $10^{-13}$, which are much below the BESII upper
limit: ${\cal BR}(J/\psi \rightarrow \bar{D}^0 e^+e^-) <1.1\times
10^{-5}$~\cite{bes2-2006}. But new physics in the loop may enhance
the decay rate considerably as discussed in Ref~\cite{zhang-2001},
further experimental search at BESIII will be interesting.

\subsection{Two-body weak hadronic decays of charmonium}

Non-leptonic two-body weak decays of the $J/\psi$ and $\psi(2S)$ are
studied based on the covariant light-front quark
model~\cite{wang-2008}, the factorization
approach~\cite{wang-2008-epj},  and
HQSS~\cite{verma-2009,lihb_twobodys_weak} in the context of the
factorization scheme for both the Cabibbo-allowed ($c \rightarrow
s$) and Cabibbo-suppressed ($c \rightarrow d$) quark-level
transitions.  In the literature, the expressions for branching
fractions for $J/\psi \rightarrow PP/PV/VV$ decays (where $P$ and
$V$ represent pseudoscalar  and vector mesons) had been given. For
$J/\psi \rightarrow PP$ mode, the dominant decay is $J/\psi
\rightarrow D^-_s \pi^+ +c.c.$ (including charge conjugate process),
which is Cabibbo and color allowed. Table~\ref{tab:psi_pp} shows the
decay rate for the $J/\psi \rightarrow PP$ modes from various
theoretical
predictions~\cite{verma-2009,wang-2008,wang-2008-epj,lihb_twobodys_weak}.

Assuming factorization as suggested by Bjorken~\cite{lihb_bjorken},
it is interesting to construct the ratio of Cabibbo suppressed to
Cabibbo allowed decay modes:
\begin{eqnarray}
\frac{{\cal BR}(J/\psi \rightarrow D_s K)}{{\cal BR}(J/\psi
\rightarrow D_s \pi)} \approx \frac{|V_{us}
f_K|^2}{|V_{ud}f_{\pi}|^2} \approx 0.081,
 \label{eq:twobodys_weak_ratio_pp}
\end{eqnarray}
where $f_K$ and $f_\pi$ are the decay constants for $K$ and $\pi$
mesons.  The ratio was predicted to be 0.08 based on the
factorization approach~\cite{wang-2008-epj}.

For $J/\psi \rightarrow PV$ mode, the Cabibbo favored and color
allowed decay $J/\psi \rightarrow D^+_s \rho^- +c.c.$ is dominant
and predicted to be $(12 \sim 50)\times 10^{-10}$ as shown in
Table~\ref{tab:psi_pv}, which can be accessible in BESIII
experiment. Moreover, the ratio, $\frac{{\cal BR}(J/\psi \rightarrow
D_s \rho)}{{\cal BR}(J/\psi \rightarrow D_s \pi)}$, was predicted to
be 6.3 in~\cite{wang-2008-epj} and 4.2 in~\cite{lihb_twobodys_weak},
which can be tested by BESIII experiment.

For $J/\psi \rightarrow PV$ mode as listed in
Table~\ref{tab:psi_vv}, the dominant decay mode is $J/\psi
\rightarrow D^{*+}_s \rho^- +c.c.$, which was predicted to be $5.3
\times 10^{-9}$ based on the the factorization
approach~\cite{wang-2008-epj}. This decay mode is the most promising
mode to be measured at BESIII.

\end{multicols}
\begin{center}
   \tabcaption{\label{tab:psi_pp}The theoretical predicted branching fractions (in unit of $
  10^{-10}$)  for $\psi \rightarrow PP$
. The transition mode, $\Delta C = \Delta S = +1$,
    corresponds to the Cabibbo-allowed decay modes, while $\Delta C = +1$,
$\Delta S = 0$ corresponds to the
    Cabibbo-suppressed decay modes. Only the central values are cited from
   the theoretical predictions. }
  \begin{tabular}{c|c|c|c|c|c}
\hline\hline
 Transition mode & Decay mode & Ref.~\cite{verma-2009} &
 Ref.~\cite{wang-2008} & Ref.~\cite{wang-2008-epj} &
 Ref.~\cite{lihb_twobodys_weak}
\\ \hline
  $\Delta C = \Delta S = +1$ & & & & & \\
                             & $\psi \rightarrow D^{-}_s\pi^{+}+c.c.$ & 7.36 &2.5 &2.0&8.74  \\
                             & $\psi \rightarrow D^0 K^0 +c.c.$ & 1.39&0.5&0.36&2.80  \\
  \hline
  $\Delta C = +1$, $\Delta S = 0$ & &  &&&\\
                                  & $\psi \rightarrow D^+_s K^- +c.c.$ & 0.53&-&0.16&0.55 \\
                                  & $\psi \rightarrow D^+ \pi^- +c.c.$ & 0.29&-&0.08&0.55 \\
                                  & $\psi \rightarrow D^0 \eta +c.c.$ & 0.070&-&-&0.016 \\
                                 & $\psi \rightarrow D^0 \eta^{\prime} +c.c.$
                                 & 0.004&-&-&0.003  \\
                                  & $\psi \rightarrow D^0 \pi^0 +c.c.$ &
                                  0.024&-&-&0.055
\\
 \hline\hline
  \end{tabular}
\end{center}
\begin{multicols}{2}

The $J/\psi$ semi-leptonic decay modes can be related to the
two-body hadronic decay modes by applying both the spin symmetry and
the non-recoil approximation to the semi-leptonic decay
rates~\cite{lihb_sl_psi}. For $J/\psi \rightarrow D^+_s( D^{*+}_s)
\pi^-$ decay modes, $q^2 = (p_{\psi} - p_{D})^2 = m_{\pi}^2$ (here
$p_{\psi}$ and $p_{D}$ are the four momenta of the initial and final
state heavy mesons) and, assuming factorization as suggested by
Bjorken~\cite{lihb_bjorken} for $B$ decays, and in the non-recoil
approximation for the hadronic transition
amplitudes~\cite{lihb_voloshin},  one can give the relation between
relative branching ratios:
\begin{eqnarray}
r&=&\frac{BR(J/\psi \rightarrow D^{*+}_s \pi^-)}{BR(J/\psi
\rightarrow D^+_s \pi^-)} \nonumber\\
&\cong& \left[\frac{d\Gamma(J/\psi \rightarrow D^{*+}_s l^- \nu
)/dq^2}{d\Gamma(J/\psi \rightarrow D^+_s l^- \nu)/dq^2}\right ]_{q^2
= m^2_{\pi}},   \label{eq:twobodys_weak_ratio_2}
\end{eqnarray}
which was predicted to be 7.5 based on the the factorization
approach~\cite{wang-2008-epj}. If a $\rho$ is substituted for the
$\pi$ one gets $r \cong 4.2$ in~\cite{wang-2008-epj}. In this way,
the estimated branching ratios in Table~\ref{tab:psi_pp} for
$\psi(nS) \rightarrow PP$ channels can be related to  $\psi(nS)
\rightarrow VP$ channels with the pseudoscalar charm mesons replaced
by vector charm mesons.
\end{multicols}
\begin{center}
   \tabcaption{\label{tab:psi_pv}
The theoretical predicted branching fractions (in unit of $
  10^{-10}$)  for $J/\psi \rightarrow PV$
. The transition mode, $\Delta C = \Delta S = +1$,
    corresponds to the Cabibbo-allowed decay modes, while $\Delta C = +1$,
$\Delta S = 0$ corresponds to
    the Cabibbo-suppressed decay modes. Only the central values are cited from
   the theoretical predictions.
}
  \begin{tabular}{c|c|c|c|c|c}
\hline\hline Transition mode & Decay mode &Ref.~\cite{verma-2009} &
 Ref.~\cite{wang-2008} & Ref.~\cite{wang-2008-epj} &
 Ref.~\cite{lihb_twobodys_weak}
\\ \hline
  $\Delta C = \Delta S = +1$ & & &&&\\
                             & $\psi \rightarrow D^+_s\rho^- +c.c.$ & 50.5&28.0&12.6&36.30 \\
                               & $\psi \rightarrow D^0 K^{*0} +c.c.$ & 8.12&5.5&1.54&10.27 \\
  \hline
  $\Delta C = +1$, $\Delta S = 0$ & &  &&&\\
                                  & $\psi \rightarrow D^+_s K^{*-}+c.c.$ &2.79&-&0.82&2.12
\\
                                  & $\psi \rightarrow D^+ \rho^- +c.c.$ &2.13&-&0.42&2.20
\\
                                 & $\psi \rightarrow D^0 \rho^0 +c.c.$ &0.18 &-&-&0.22 \\
                                  & $\psi \rightarrow D^0 \omega +c.c.$ & 0.16&-&-&0.18
\\
                                  & $\psi \rightarrow D^0 \phi +c.c.$ &0.41&-&-&0.65 \\
 \hline\hline
  \end{tabular}
\end{center}
\begin{multicols}{2}

\begin{center}
   \tabcaption{\label{tab:psi_vv}
The theoretical predicted branching fractions (in unit of $
  10^{-10}$)  for $J/\psi \rightarrow VV$
.  Only the central values are cited from
   the theoretical predictions.}
  \begin{tabular}{c|c}
\hline\hline Decay mode &  Ref.~\cite{wang-2008-epj}
\\ \hline
 $J/\psi \rightarrow D^{*+}_s\rho^- +c.c.$ & 52.6 \\
 $J/\psi \rightarrow D^{*+}_s K^{*-} +c.c.$ & 2.6 \\
 $J/\psi \rightarrow D^{*+} \rho^- +c.c.$ & 2.8 \\
 $J/\psi \rightarrow D^{*+} K^{*-} +c.c.$ & 9.6\\

 \hline\hline
  \end{tabular}
\end{center}

All of the above estimates show an overall enhancement of the
final-state vector charm mesons with respect to the pseudoscalar
ones.  This suggests the use of $D^*_s$ or $D^{*\pm}$ as signals in
searches for weak decays of the $J/\psi$ in non-leptonic decay
channels as well. According to the predicted results in
Ref.~\cite{wang-2008-epj}, the branching fraction for inclusive weak
hadronic decays of $J/\psi$ can be as large as $1.3\times 10^{-8}$,
which is in remarkable agreement with the naive estimation in
Ref.\cite{lihb_sl_psi}.

In the SM, FCNC induced $J/\psi$ weak hadronic decays are predicted
to be unobservably small~\cite{lihb_weak} and, thus, any observation
of such decay would provide a signal for new physics. In
Ref.~\cite{lihb_weak} the predictions of various models, such as
TopColor models, minimal supersymmetric standard model (MSSM) with
R-parity violation and a general two-Higgs-doublet model, are
discussed. These authors found that the branching fraction for
$J/\psi \rightarrow D/\bar{D} X_u$, which is mediated by the weak $c
\rightarrow u$ transition, could be as large as $10^{-6}\sim
10^{-5}$ in some new physics scenarios.

At BESIII  it will be difficult to isolate pure, $c \rightarrow u$
mediated, hadronic $J/\psi \rightarrow D^{(*)}/\bar{D}^{(*)} X_u$
decays. On the other hand the decays $J/\psi \rightarrow
D^{0(*)}/\bar{D}^{0(*)} l^+ l^-$ ($l= e$, $\mu$) and $J/\psi
\rightarrow D^{0(*)}/\bar{D}^{0(*)} \gamma $ decays, which are also
dominated by FCNC processes, would be quite distinct.

In addition to the $J/\psi$ and $\psi(2S)$ decays into charm meson,
the weak decays of $\psi(2S)$ into final states involved $\Lambda_c$
baryon can also be searched for at BESIII. These decay modes include
$\psi(2S) \rightarrow \Lambda_c^+ \bar{\Sigma}^- +c.c.$,
$\Lambda_c^+ \bar{\Sigma}^0 \pi^- +c.c.$, $\Lambda_c^+
\bar{\Sigma}^- e^- \bar{\nu} +c.c.$ and $\Lambda_c^+
\bar{\Lambda}\pi^- +c.c.$. In ref.~\cite{li2011}, the decay of
$\psi(2S) \rightarrow \Lambda_c^+ \bar{\Sigma}^- +c.c.$ are
estimated in the framework of the SM, and they found that the decay
rate could reach $10^{-10}$.

\subsection{Searches at BESIII}
\label{sec:cc-weak-decays}

 At BESIII, assuming a $10^{10}$ $J/\psi$
event sample, the total semi-leptonic decay rate could be $ 5 \times
10^{-9}$, so about 50 semi-leptonic decay events of the type $J/\psi
\rightarrow D_s (D^*_s) l \nu$ can be accumulated by BESII per year.
The following event selection criteria would be useful for searching
 such exclusive semi-leptonic channels :
\begin{itemize}
 \item The prompt charged lepton can be used to tag the weak decay:
in order to suppress cascade decay backgrounds from $J/\psi$ strong
decays, the tagging lepton momentum could be required to be between
0.5 GeV and 1.0 GeV, close to the upper kinematic limit for the
decay under consideration.  High quality lepton discrimination from
charged pions or kaons is needed for the measurement.
 \item The missing mass of the reconstructed candidates must be consistent
with the (nearly) zero mass of the undetected neutrino.
 \item The reconstruction of a $D_s$ or $D^{\pm}$ meson would provide
an unambiguous signature for a weak decay of a below-open-charm
threshold $\psi(nS)$.  Good invariant mass resolution of the ${D_s}$
decay products will be important for reducing combinatorial
backgrounds.
 \item Soft photons in the energy interval (90$\sim$200)~MeV
from the $D^*_s\rightarrow \gamma D_s$ transition and soft charged
pions from $D^{*\pm}\rightarrow \pi^{\pm}D$ decay can provide
further suppression of combinatorial backgrounds: the additional
constraint of an intermediate $D^*_s$ state would reconfirm the
$D_s$ signal.
\end{itemize}

 In general, exclusive hadronic decays are probably too
tiny to look for in any specific fully reconstructed decay channel.
Therefore, it seems that an inclusive search for $J/\psi \rightarrow
D^*_s + X$ at BESIII may be more fruitful. The $\gamma$ from the
decay of a $D^*_s$ meson should be useful as a kinematic constraint
to clean up any $D_s$ meson signal, as discussed
in~\cite{lihb_sl_psi}.

\section{Search for the invisible decays of quarkonium}
\label{sec:invisible_cc_decays}

Invisible decays of quarkonium states offer a window to look for
possible new physics beyond the
SM~\cite{lihb_Part4_Fayet:1979qi,lihb_Part4_Fayet:2006sp}. The
reason is that other than neutrinos, the SM includes no other
invisible final particles that these states can decay into.  BESII
explored such a window by establishing the first experimental limits
on invisible decays of the $\eta$ and $\eta'$, which complemented
the limit of $2.7 \times 10^{-7}$ that was previously established
for invisible  decays of the
$\,\pi^\circ$~\cite{lihb_Part4_Artamonov:2005cu}.

Some theories of beyond the SM physics predict new particles with
masses that are accessible at BESIII, such as the light dark matter
(LDM) particles discussed in Ref.~\cite{lihb_Part4_Boehm:2003hm}.
These can have the right relic abundance to constitute the
nonbaryonic dark matter of the universe, if they are coupled to the
SM via a new light gauge boson $U$~\cite{lihb_Part4_Fayet:1980ad},
or exchanges of heavy fermions.  A light neutralino with a coupling
to the SM that is mediated by a light scalar singlet in the
next-to-minimal supersymmetric standard model has also been
considered~\cite{lihb_Part4_Ellis:1988er}.

These considerations have received a boost in interest by the recent
observation of a bright~511 keV $\gamma$-ray line from the galactic
bulge reported by the SPI spectrometer on the INTEGRAL
satellite~\cite{lihb_Part4_Jean:2003ci}.  The corresponding galactic
positron flux, as well as the smooth symmetric morphology of the
511~keV $\gamma$ emission, could be interpreted as originating from
the annihilation of LDM particles into $e^+e^-$ pairs
~\cite{lihb_Part4_Boehm:2003hm,lihb_Part4_Beacom:2004pe}. It would
be very interesting to see evidence for such light invisible
particles in collider experiments.  CLEO gave an upper bound on
$\Upsilon(1S) \rightarrow \gamma + \mbox{invisible}$, which is
sensitive to dark matter candidates lighter than about 3
GeV/$c^2$~\cite{lihb_Part4_Balest:1994ch}, and also provided an
upper limit on the axial coupling of any new $U$ boson to the $b$
quark.  It is important, in addition, to search for the invisible
decays of othr light quarkonium states ($q\bar{q}$, $q= u$,$d$, or
$s$ quark), since these can be used to constrain the masses of LDM
particles and the couplings of a $U$ boson to the light
quarks~\cite{lihb_Part4_Fayet:2006sp}.

 It has been shown
that measurements of the $J/\psi$ invisible decay widths can be used
to constrain new physics models~\cite{lihb_Part4_ng_invisible}. It
is straightforward for one to calculate the SM ratio of branching
fractions for $J/\psi$  invisible decays and its measured branching
fraction for decays into electron-positron
pairs~\cite{lihb_Part4_ng_invisible}. Within the SM, the invisible
mode consists solely of decays into the three types of
neutrino-antineutrino pairs. Neglecting polarization effects and
taking into account $e^+e^-$ production through a photon only, one
gets~\cite{lihb_Part4_ng_invisible}:
\begin{eqnarray}
\frac{\Gamma(J/\psi \rightarrow \nu\bar{\nu})}{\Gamma(J/\psi
\rightarrow e^+e^-)} &=& \frac{27 G^2 M^4_{J/\psi}}{256
\pi^2\alpha^2}
\left(1-\frac{8}{3}sin^2(\theta_W \right)^2 \nonumber \\
&=& 4.54 \times 10^{-7}, \label{eq:invisible_ratio}
\end{eqnarray}
where $G$ and $\alpha$ are the Fermi and fine-structure constants,
respectively, and $M_{J/\psi}$ is the $J/\psi$ mass. The uncertainty
of the above relation is about 2$\sim$3\% and comes mainly from the
corrections to the $J/\psi$ wave function, $e^+e^-$ production via
the $Z$ boson and electroweak radiative
corrections~\cite{lihb_Part4_ng_invisible}.

In BES experiment, one can tag the charmonium states that decay
invisibly by looking for a particular cascade transition, such as
$\psi(2S) \rightarrow \pi^+\pi^- J/\psi$, $\psi(2S) \rightarrow
\gamma \chi_c$ and so on, where the soft $\pi^+\pi^-$ pair or the
monoenergetic radiative $\gamma$ serves as a tag for the invisibly
decaying $J/\psi$ or $\chi_c$ state. The BESII experiment performed
the first search for invisible decays of the $J/\psi$ using
$\psi^\prime \rightarrow \pi^+\pi^- J/\psi$ events detected in a
sample of 14.0 million $\psi^\prime$ decays. The upper limit on the
ratio $\frac{{\cal B}(J/\psi\rightarrow \mbox{invisible})}{{\cal
B}(J/\psi\rightarrow \mu^+\mu^-)}$ at the 90\% confidence level is
$1.2\times 10^{-2}$~\cite{prl-psi-invi}. This measurement improves
by a factor of 3.5 the bound on the product of the coupling of the
$U$ boson to the $c$ quark and LDM particles as described in Eqs.
(25) and (26) of Ref.~\cite{00}. One now has, for a Majorana LDM
particle $\chi$ as in Eq. (26) of Ref.~\cite{00}, a limit of
$|c_{\chi}f_{cV}|< 8.5\times 10^{-3}$, which is almost a factor of 2
stronger than the corresponding limit $|c_{\chi}f_{bV}|<1.4\times
10^{-2}$ derived from the invisible decays of the $\Upsilon(1S)$ as
described in Eq. (106) in Ref.~\cite{000}, where $c_{\chi}$ and
$f_{cV}$ ($f_{bV}$) denote the $U$ boson couplings to the LDM
particle $\chi$ and $c$ ($b$) quark. We expect a more precise
measurement can be obtained in the future BES-III experiment. A list
of potentially useful decay chains is provided in
Table~\ref{tab:invisible_mode_cc} in the BESIII experiment.

\end{multicols}

\begin{center}
  \tabcaption{\label{tab:invisible_mode_cc} $\psi(2S)$ and $J/\psi$ decay modes that can be used
to search for invisible
           decays of the $J/\psi$, $\chi_{c0}$, $\chi_{c1}$,
$\chi_{c2}$, $\eta_c(1S)$ and $\eta_c(2S)$.
            The branching fractions are taken from the
PDG~\cite{lpv_PDG}. For each mode, a
            ``tagging topology" is given, which is the set of visible
particles that are seen within the detector's acceptance. In
            each case the tagging topology has well defined kinematics.
           The number of events are the expected event yield in 3
billion $\psi(2S)$ (10 billion $J/\psi$) data set, in which we did
not consider the decay probabilities of
            the tagging particles.  }
  \footnotesize
  \begin{tabular}{c|c|c|c|c}\hline
\hline $\psi(2S)$ & Branching  &Number of events  & Invisible &
Tagging
 \\
    decay mode & fraction ($10^{-2}$) & expected
 & decay mode & topology  \\\hline

   $\psi(2S) \rightarrow \pi^+ \pi^- J/\psi$ & $31.7 \pm 1.1$ & $9.3\times
10^8$ & $J/\psi \rightarrow \mbox{invisible}$
    & $\pi^+ \pi^-$  \\ \hline
  $\psi(2S) \rightarrow \pi^0 \pi^0 J/\psi$ & $18.6 \pm 0.8$& $5.6\times
10^8$ & $J/\psi \rightarrow  \mbox{invisible}$
    & $\pi^0 \pi^0$  \\ \hline
   $\psi(2S) \rightarrow \eta J/\psi$ & $3.08 \pm 0.17$ &$9.3
\times 10^7$ & $J/\psi \rightarrow  \mbox{invisible}$
    & $\eta$ \\ \hline
  $\psi(2S) \rightarrow \pi^0 J/\psi$ & $0.123 \pm 0.018$ &$3.7
\times 10^6$ & $J/\psi \rightarrow  \mbox{invisible}$
    & $\pi^0$  \\ \hline
  $\psi(2S) \rightarrow \gamma \chi_{c0}$ & $9.0 \pm
0.4$ &$2.7\times 10^8$ & $\chi_{c0}  \rightarrow  \mbox{invisible}$
    & $\gamma$ \\ \hline
  $\psi(2S) \rightarrow \gamma \chi_{c1}$ & $8.7 \pm 0.5$ &$2.6
\times 10^8$ & $\chi_{c1}  \rightarrow  \mbox{invisible}$
    & $\gamma$  \\ \hline
 $\psi(2S) \rightarrow \gamma \chi_{c2}$ & $8.2 \pm 0.3$ &$2.5
\times 10^8$ & $\chi_{c2} \rightarrow  \mbox{invisible}$
    & $\gamma$  \\ \hline
 $\psi(2S) \rightarrow
\gamma \eta_c(1S)$ & $0.26 \pm 0.04$ &$7.8\times 10^6$ & $\eta_c(1S)
\rightarrow  \mbox{invisible}$
    & $\gamma$  \\ \hline
 $J/\psi \rightarrow \gamma \eta_c(1S)$ & $1.3 \pm 0.4$ & $1.3
\times 10^8$ & $\eta_c(1S) \rightarrow  \mbox{invisible}$
    & $\gamma$ \\ \hline\hline
  \end{tabular}
\end{center}

\begin{multicols}{2}

It is also interesting to search for invisible decays of the $\eta$,
$\eta^{\prime}$,  $\rho$, $\omega$ and $\phi$ light mesons. Within
the SM, the decays of $\eta(\eta{'})\to\nu\bar{\nu}$ are tiny due to
the helicity suppression. If the $Z^0$ couples to a massive neutrino
with the standard weak interaction strength, the branching ratios
(BR) for $\pi^0 \rightarrow \nu_{\tau} \bar{\nu_{\tau}}$ and $\eta
\rightarrow \nu_{\tau} \bar{\nu_{\tau}}$ have maximum values of
$5.0\times 10^{-10}$ and $1.3\times
10^{-11}$~\cite{smpredict,cubic}, respectively, at the $\nu_{\tau}$
mass upper limit of $m_{\nu_{\tau}} =18.2$ MeV/c$^2$~\cite{lpv_PDG}.
Any enhanced signal of invisible decay may indicate New Physics.
Possible $\eta/\eta^\prime \rightarrow$invisible decay products
could be LDM particles or light neutralinos. These LDM particles may
have an adequate relic density to account for the non-baryonic mass
of the universe.

In the BES experiment, the two-body decay modes $J/\psi \rightarrow
\phi\eta$ or $\phi\eta^{\prime}$ can be selected using only the very
clean and distinct $\phi\rightarrow K^+K^-$ decays, which then tag
the presence of an $\eta$ or $\eta^{\prime}$ meson that has decayed
into an invisible final state. For $J/\psi\to\phi\eta^\prime$, the
missing momentum, ${P}_{miss} = |\vec{P}_{miss}|$, is a powerful
discriminating variable to separate signal events from possible
backgrounds in which the missing side is not from an $\eta$
($\eta^{\prime}$). Here, $\vec{P}_{miss} = - \vec{P}_{\phi}$. In
addition, the regions of the detector where the $\eta$ and
$\eta^{\prime}$ decay products are expected to go are easily defined
thanks to the strong boost from the $J/\psi$ decay.

The branching fraction of $\eta (\eta^{\prime})\rightarrow \gamma
\gamma$ was also determined in $J/\psi \rightarrow \phi \eta
(\eta^{\prime})$ decays, in order to provide the ratio
$\mathcal{B}(\eta(\eta^\prime) \rightarrow {\rm invisible})$ to
$\mathcal{B}(\eta(\eta^\prime) \rightarrow \gamma \gamma)$.  The
advantage of measuring
$\displaystyle\frac{\mathcal{B}(\eta(\eta^\prime) \rightarrow
  {\rm invisible})}{\mathcal{B}(\eta(\eta^\prime) \rightarrow \gamma
  \gamma)}$ is that the uncertainties due to the total number of
$J/\psi$ events, tracking efficiency, PID, the number of the charged
tracks, the cut on $M(KK)$, and residual noise in the BSC all
cancel.

Based on $58\times 10^6$ $J/\psi$ events in the BESII experiment,
the upper limit on the ratio of the $\mathcal{B}(\eta\to {\rm
invisible})$ to $\mathcal{B}(\eta\to\gamma\gamma)$ was determined to
be~\cite{lihb_bes-ii-invisible}:
\begin{eqnarray}
  \frac{\mathcal{B}(\eta(\eta^\prime) \to
{\rm invisible})}{\mathcal{B}(\eta (\eta^\prime) \to\gamma\gamma)} <
1.65\times 10^{-3} (6.69\times 10^{-2}).
  \label{eq:upper}
\end{eqnarray}

Table~\ref{tab:invisible_lighter} lists the possible two-body
$J/\psi$ decay modes that can be used to study the invisible decays
of the $\eta$, $\eta^{\prime}$, $\rho$, $\omega$ and $\phi$ mesons
at BESIII.

\end{multicols}
\vspace{2cm}
\begin{center}
  \tabcaption{\label{tab:invisible_lighter} $J/\psi$ decay modes that can be used to study invisible
decays of $\eta$, $\eta^{\prime}$, $\rho$, $\omega$ and $\phi$
mesons.
 The branching fractions are from the PDG~\cite{lpv_PDG}. For
each mode, a ``tagging topology" is given, which is the set of
visible tracks in the detector's acceptance. In each case the
tagging topology has well defined kinematics.  The produced number
of events are the expected events in 10 billion $J/\psi$ event data
set at BESIII, with the decay probabilites of the tagging particles
included. }
  \footnotesize
  \begin{tabular}{c|c|c|c|c}\hline
    $J/\psi$ & Branching & Invisible & Tagging & Number of events \\
    decay mode & fraction ($10^{-4}$)  & decay mode & topology & /10
billion $J/\psi$ decays \\\hline
   $J/\psi \rightarrow \phi \eta $ & $6.5 \pm 0.7$ &  $\eta \rightarrow
     \mbox{invisible}$
    & $\phi \rightarrow K^+ K^-$ & $(31.4 \pm 3.4)\times 10^{5}$\\
      & $6.5 \pm 0.7$ & $\phi \rightarrow  \mbox{invisible}$
    & $\eta \rightarrow \gamma \gamma $ &  $(25.7 \pm 2.8 )\times 10^{5}$
\\\hline
      $J/\psi \rightarrow \phi \eta^{\prime} $ & $3.3 \pm 0.4$ &
$\eta^{\prime} \rightarrow  \mbox{invisible}$
    & $\phi \rightarrow K^+ K^- $ &  $(16.2 \pm 1.9 )\times 10^{5}$ \\
         & $3.3 \pm 0.4$ & $\phi \rightarrow  \mbox{invisible}$
    & $\eta^{\prime} \rightarrow \gamma \rho^0 $ &
$(9.6\pm 1.2 )\times 10^{5}$ \\\hline

      $J/\psi \rightarrow \omega \eta $ & $15.8 \pm
1.6$ & $\eta \rightarrow  \mbox{invisible}$
    & $\omega \rightarrow \pi^+\pi^-\pi^0 $ &
$(13.9 \pm 1.4 )\times 10^{6}$ \\
       & $15.8 \pm 1.6$ & $\omega \rightarrow  \mbox{invisible}$
    & $\eta \rightarrow \gamma \gamma $ &
$(6.2 \pm 0.6 )\times 10^{6}$\\ \hline

      $J/\psi \rightarrow \omega \eta^{\prime} $ & $1.67 \pm 0.25$ &
$\eta^{\prime} \rightarrow  \mbox{invisible}$
    & $\omega \rightarrow  \pi^+\pi^-\pi^0 $ & $(1.5 \pm 0.2)\times
10^{6}$\\
        & $1.67 \pm 0.25$ & $\omega \rightarrow  \mbox{invisible}$
    & $\eta^{\prime} \rightarrow \gamma \rho^0 $ & $(0.7 \pm 0.1 )\times
10^{6}$\\ \hline

      $J/\psi \rightarrow \rho^0
\eta $ & $1.93 \pm 0.23$ & $\eta \rightarrow  \mbox{invisible}$
    & $\rho^0 \rightarrow \pi^+\pi^- $ &  $(1.9 \pm 0.2 )\times 10^{6}$\\
      & $1.93 \pm 0.23$ & $\rho^0 \rightarrow  \mbox{invisible}$
    & $\eta \rightarrow \gamma \gamma $ & $(0.8 \pm 0.09 )\times 10^{6}$
\\
\hline

      $J/\psi \rightarrow \rho^0 \pi^0 $ & $56 \pm
7$ & $\rho^0 \rightarrow  \mbox{invisible}$
    & $\pi^0 \rightarrow \gamma\gamma $ &   $(55.3 \pm 5.8 )\times
10^{6}$\\
\hline

  \end{tabular}
\end{center}
\begin{multicols}{2}

\section{Search for $C$ or $P$ violating processes in $J/\psi$ decays}
\label{sec:c_cc_decays}

With its huge $J/\psi$ and $\psi(2S)$ data samples, the BESIII
experiment will be approaching the statistics regime where studies
of rare $\psi$ decays can provide important tests of the SM and
possibly uncover deviations. Among the interesting examples are $C$,
$P$ or $CP$ violating processes in $J/\psi$ decays.  An example of
such modes would be $\psi(nS) \rightarrow V^0 V^0$, where $V^0$ is
used to denote $J^{PC} = 1^{--}$ vector mesons ($\phi$, $\omega$,
$\rho^0$ and $\gamma$).  A distinct signal for this class of event
would be $\psi(nS) \rightarrow \phi \phi$  detected in $\psi(nS)
\rightarrow K^+K^- K^+K^-$ final states.  Because of the $C$
violation,  $\psi(nS)\rightarrow V^0 V^0$ decays can only occur in
the SM via $c\bar{c}$ annihilation via a $Z^0$ or $W$-exchange
decays as discussed in Ref.~\cite{lihb_double_phi}. The rate for
this type of weak decay can provide a measurement of the charmonium
wave function at the origin~\cite{lihb_double_phi}.

In order to make a rough estimate for the rate, we first consider
just the rate due to the $W$-exchange contribution, which is
straightforward to compute~\cite{lihb_goggi_cp}
\begin{eqnarray}
\frac{\Gamma(J/\psi \rightarrow s\bar{s})^{weak}} {\Gamma (J/\psi
\rightarrow e^+e^-)} \cong \frac{1}{2} \left( \frac{m_{J/\psi}}{m_W}
\right)^4, \label{eq:cp_vv}
\end{eqnarray}
where $m_{J/\psi}$ and $m_{W}$ are the masses of $J/\psi$ and $W$
boson,  respectively.  This leads to $BR(J/\psi \rightarrow
s\bar{s})^{weak} \cong 10^{-7}$ for this weak contribution. To form
the $\phi\phi$ final state, another $s\bar{s}$ pair must be produced
from the vacuum and these $s$-quarks have to bind with the outgoing
$s\bar{s}$ from the $c\bar{c}$ decay to produce the $ \phi \phi$
final state~\cite{lihb_double_phi}.
When this is considered, it seems that one can expect that the SM
exclusive $BR(\psi(nS) \rightarrow \phi \phi)$ rate should be below
the level of $10^{-8}$ and probably out of reach of the BESIII
experiment.

Experimentally, there are some possible backgrounds that will dilute
the signal for $J/\psi \rightarrow \phi\phi$ decays. One major
background is $J/\psi \rightarrow \gamma \phi \phi$, which is mainly
from $J/\psi \rightarrow \gamma \eta_c(1S)$, $\eta_c(1S) \rightarrow
\phi \phi$. This background can be  removed by doing a constrained
kinematic fit. A detailed calculation had been done to estimate the
background from $J/\psi \rightarrow \gamma \phi
\phi$~\cite{lihb_double_phi}. Another background appears if one
studies only $2(K\bar{K})$ invariant pair mass distributions. It
arises from the $C$ and $P$-conserving reaction $J/\psi \rightarrow
\phi (K\bar{K})_{S-wave}$, due to the fact that the $\phi$ mass is
only two $S$-wave-widths away from the $K\bar{K}$~$S$-wave resonance
mass, for example, $f_0(980) \rightarrow K\bar{K}$. Although it may
be difficult to subtract in a small statistical sample, one can, in
principle, remove this kind of background by either a spin-parity
analysis of the $ K\bar{K}$ pairs in a narrow window about $\phi$
mass, or by a subtraction normalized to an observed $S$-wave mass
peak.  To avoid the $S$-wave contribution, one can reconstruct one
$\phi$ from $K^+K^-$ and another $\phi$ from the $K_S K_L$ mode,
which is not allowed to form an $S$-wave. It will be easy to look
for the missing mass of one $\phi$ reconstructed from $K^+K^-$, to
see if there is any peak under the $\phi$ mass region by also
requiring $K_S$ and $K_L$ information in the final states.


 It is noted that there is possible continuum background
produced via a two-photon annihilation process. It is a peaking
background that cannot be removed without considering detailed
angular distributions in a high statistics sample.  It is very hard
to deal with this kind of peaking source with a small sample of
signal events. One way is to use off-peak data which are taken below
the $J/\psi$ mass peak. The $e^+e^- \rightarrow \gamma \gamma$
process has been investigated before~\cite{lihb_two_photons}, and it
has a unique production angle ($\theta^*$) distribution, which is
defined as the angle between $\phi$ and $e^-$ beam direction in the
Center-of-Mass (CM) frame. The production angle distribution for the
two real photon annihilation process has the form of
\begin{eqnarray}
 \sigma (cos\theta^*)_{e^+e^- \rightarrow \gamma \gamma} =
\frac{1+cos\theta^{*2}}{1-cos\theta^{*2}},
 \label{eq:angular_twophoton}
\end{eqnarray}
while, in the process of two virtual photon into $V^0V^0$ pairs, the
distribution is (to first order)~\cite{lihb_stan}:
\begin{eqnarray}
 \sigma (cos\theta^*)_{e^+e^- \rightarrow \gamma^*
\gamma^* \rightarrow V^0V^0} =
\frac{1+cos\theta^{*2}}{k^2-cos\theta^{*2}},
 \label{eq:angular_twophoton_1}
\end{eqnarray}
where factor $k$ is:
\begin{eqnarray}
 k= \frac{2m^2_{V^0}-S}{\sqrt{S^2 - 4Sm^2_{V^0} }},
 \label{eq:angular_twophoton_2}
\end{eqnarray}
where $S$ is the square of CM energy. In principle, by using an
angular analysis, one can remove the peaking background with high
statistic data sample. To avoid the peaking background from the
continuum, $\psi(2S) \rightarrow \pi \pi J/\psi$ could be used to
study this kind of rare $J/\psi$ decays with 3 billion $\psi(2S)$
sample, but the statistics will be substantially reduced.

\section{Charged-lepton flavor violating processes in decays of $J/\psi$ }
\label{sec:lfv_cc_decays}

In the framework of the SM, lepton flavor is an accidentally
conserved quantum number due to the massless neutrinos. The new
physics beyond the SM introduces oscillations between the neutrino
flavors by including the neutrino masses, which violate lepton
flavor. However, SM processes involving charged-lepton flavor
violation (CLFV) with massive neutrinos are too tiny to be observed
because they are suppressed by the quantity $(\Delta
M^2_{\nu}/M^2_{W})^2 <10^{-48}$~\cite{strumia-2006}. Here $\Delta
M_{\nu}$ is the difference between the masses of neutrinos of
different flavor and $M_W$ is the mass of the charged weak vector
boson. Hence, CLFV offers an opportunity to probe signature of new
physics~\cite{ellis-2000,ellis-2004,salam-1974,glashow-1974,li-wei-2009}.
The CLFV charmonium decays $J/\psi \rightarrow l l^{\prime}$ ($l$
and $l^{\prime} = \tau, \mu, e$, $l \ne l^{\prime}$) are predicted
by various theoretical models that allow tree-level FCNC, including,
e.g., unparticle theory~\cite{li-wei-2009,lu-wang-2009}, $R$-parity
violating and large tan$\beta$ SUSY scenarios, leptoquarks, and
other models inspired by the idea of grand
unification~\cite{salam-1974,glashow-1974}.

By using unitarity considerations, limits on CLFV $\mu$ and $\tau$
branching fractions~\cite{lpv_PDG} have been used to place indirect
limits on CLFV $J/\psi$ branching
fractions~\cite{lpv_zhang,lpv_zhang_3,gut2011}. In
Ref.~\cite{lpv_zhang}, unitarity relations between the CLFV $J/\psi$
decay and the CLFV $\mu$ and $\tau$ decays have been exploited. From
the existing experimental bounds on the latter
process~\cite{lpv_PDG}, the stringent indirect limits can be
obtained to be $4\times 10^{-13}$ and $6 \times 10^{-9}$ for $J/\psi
\rightarrow \mu e$ and $\tau \mu/\tau e$, respectively. In
Ref.~\cite{gut2011}, the experimental bounds on nuclear $\mu^- -
e^-$ conversion are used to place indirect limits on the CLFV
$J/\psi$ decays, which are at the same level as those in
Ref.~\cite{lpv_zhang}. As suggested by Zhang in
Ref.~\cite{lpv_zhang_3}, searching for the CLFV $J/\psi$ decays with
a huge sample at the BESIII remains a worthwhile experimental
challenge. With a 58 M $J/\psi$ event sample at BESII, the following
upper limits have been established~\cite{lpv_besii}:
\begin{eqnarray}
{\cal BR}(J/\psi \rightarrow \tau^{\pm} e^{\mp}) < 8.3 \times 10^{-6}; \\
{\cal BR}(J/\psi \rightarrow \tau^{\pm} \mu^{\mp}) < 2.0 \times 10^{-6}; \\
{\cal BR}(J/\psi \rightarrow \mu^{\pm} e^{\mp}) < 1.1 \times
10^{-6}. \label{eq:lpv_psi_besii}
\end{eqnarray}
The limits on the two-body  lepton flavor violating decays of the
$J/\psi$ could be reduced to the $10^{-8} \sim 10^{-9}$ level at
BESIII with one year full-luminosity run at the $J/\psi$ peak. This
would be a significant improvement.

We should note that the indirect limits on CLFV $J/\psi$ decays are
significantly more stringent than the expected bounds from the
BESIII experiment. However, searches for the CLFV decays of vector
mesons remain an important experimental effort since their
observation at the rates above the indirect limits  would be a
manifestation of new CLFV physics, which does not fit into these
theoretical analyses. In particular, it may imply a nontrivial
mechanism of self cancellation as discussed in
Refs.~\cite{lpv_zhang,lpv_zhang_3,gut2011}, which are considered as
unnatural.

\section{Summary}

A mini-review of $J/\psi$ and $\psi(2S)$ rare decays in the BES-III
experiment has been done. With one year's luminosity data, about 10
billion $J/\psi$ and 3 billion $\psi(2S)$ decays can be collected at
BESIII experiment. The searches of weak decay, LFV decay, $C$ or $P$
violated decay and invisible decays of charmonium states will be
feasible. These rare decay processes will provide opportunity for
new physics space, especially, in charm sector.  It should be noted
that the recent measurement of 3.5$\sigma$ evidence for direct $CP$
violation in $D^0$ decay at LHCb may indicate possible contributions
to charm decays from physics beyond SM. We expect to see surprises
from BESIII experiment in rare charmonium decays.

\acknowledgments{The authors would like to thank X.~M.~Zhang and
M.~Z.~Yang for useful discussions. }

\end{multicols}

\vspace{10mm}

\vspace{-1mm}
\centerline{\rule{80mm}{0.1pt}}
\vspace{2mm}

\begin{multicols}{2}

\end{multicols}

\clearpage

\end{document}